# Structure and mechanical properties of monolayer amorphous carbon and boron nitride

Xi Zhang[1#], Yu-Tian Zhang[1,2#], Yun-Peng Wang[3], Shiyu Li[1], Shixuan Du[1,4], Yu-Yang Zhang[1], Sokrates T. Pantelides[5,1]

1. University of Chinese Academy of Sciences and Institute of Physics, Chinese Academy of Sciences, Beijing 100049, China

2. CAS Key Laboratory of Theoretical Physics, Institute of Theoretical Physics, Chinese Academy of Sciences, Beijing 100190, China

3. Hunan Key Laboratory for Super Microstructure and Ultrafast Process, School of Physics and Electronics, Central South University, Changsha 410083, China

4. Songshan Lake Materials Laboratory, Dongguan, Guangdong 523808, China

5. Department of Physics and Astronomy and Department of Electrical and Computer Engineering, Vanderbilt University, Nashville, Tennessee 37235, USA

**Abstract**

Amorphous materials exhibit various characteristics that are not featured by crystals and can sometimes be tuned by their degree of disorder (DOD). Here, we report results on the mechanical properties of monolayer amorphous carbon (MAC) and monolayer amorphous boron nitride (maBN) with different DOD. The pertinent structures are obtained by kinetic-Monte-Carlo (kMC) simulations using machine-learning potentials (MLP) with density-functional-theory (DFT)-level accuracy. An intuitive order parameter, namely the areal fraction $F_x$ occupied by crystallites within the continuous random network, is proposed to describe the DOD. We find that $F_x$ captures the essence of the DOD: Samples with the same $F_x$ but different sizes and distributions of crystallites have virtually identical radial distributions functions as well as bond-length and bond-angle distributions. Furthermore, by simulating the fracture process with molecular dynamics, we found that the mechanical responses of MAC and maBN before fracture are solely determined by $F_x$ and are insensitive to the sizes and specific arrangements of the crystallites. The behavior of cracks in the two materials is analyzed and found to mainly propagate in meandering paths in the CRN region and to be influenced by crystallites in distinct ways that toughen the material. The present results reveal the relation between structure and mechanical properties in amorphous monolayers and may provide a universal toughening strategy for 2D materials.

**Introduction**

Two-dimensional (2D) materials exhibit unique properties. Their mechanical properties, in particular fracture toughness, which describes the ability of a material containing a crack to resist fracture, are essential for their reliable integration into future electronic, composite, and nano-electromechanical applications [1-4]. However, cracks in 2D materials generally induce brittle behavior at room temperature [5-8]. Given the brittle nature of 2D materials, it is important to investigate their mechanical properties and find effective ways to toughen them for applications. Introducing extrinsic defects and increasing the defect density is one way to increase the fracture toughness of graphene [9]. In contrast, binary materials like monolayer h-BN are intrinsically toughened by an asymmetric deformation at crack tips (due to asymmetric edge polarization) [10]. Overall, disorder engineering is an effective toughening strategy for 2D materials.

Amorphous materials that are highly disordered feature a wealth of mechanical properties [11-16], but their atomic structures are very complicated and highly debated. As a result, the construction of structure-properties relations for amorphous materials remains a long-standing riddle. The task is simpler in 2D, as it is possible to directly determine the atomic positions by high-resolution scanning transmission electron microscopy (STEM). In 2019, monolayer amorphous carbon (MAC) was successfully synthesized for the first time and atomic-resolution STEM directly revealed that MAC is a Zachariasen continuous random network (Z-CRN) containing crystallites. It was also found that MAC exhibits high toughness [17]. More recently, in the case of MAC, the degree of disorder (DOD) was found to be tunable by the growth temperature and to affect the electrical conductivity significantly [18]. Two order parameters that can be measured experimentally were introduced to correlate properties to the DOD.

Monolayer amorphous BN (maBN) has not been synthesized so far (only amorphous thin films have been reported [19]). The structure of maBN has been studied by kinetic Monte Carlo (kMC) simulations using empirical potentials [20]. It was found that maBN features pseudocrystallites, i.e., honeycomb regions comprising noncanonical hexagons with random B-B and N-N bonds, in a Z-CRN [20]. Furthermore, the mechanical and thermal properties of MAC and maBN have by now also been investigated by simulations based on empirical potentials [20-23]. However, the accuracy of empirical potentials is never as high as high as that of DFT calculations, especially for binary materials. kMC simulations based on DFT evaluations of total energies for the construction of amorphous structures still remain out of reach, but the advent of practical methods for generating DFT-based machine-learning potentials opens up new opportunities to investigate the structure and properties of amorphous materials.

In this paper, we investigate the structure and mechanical properties of monolayer amorphous carbon and boron nitride using machine learning potentials (MLP) with density-functional theory (DFT)-level accuracy. The kMC simulation [24], a widely-used sampling method for fast exploration of potential energy surfaces, was employed to assist the active-learning procedure to train the MLPs. Then the structure evolution of MAC and maBN is simulated by kMC with the bonding energetics described by the as-trained MLPs. It is found that crystallites are in fact more energetically favored within maBN than pseudocrystallites. Moreover, an intuitive order parameter, $F_x$, the fraction of the area occupied by crystallites, is proposed to quantify the DOD of these amorphous materials. We find that $F_x$ captures the essence of the DOD: We demonstrate that *samples with very different atomic structures but the same $F_x$ have essentially identical radial distribution functions and bond-angle and bond-length distributions*. As a result, the mechanical properties of MAC and maBN samples with different DOD, namely the critical stress and strain that lead to fracture, investigated by MLP-based molecular dynamics (MD) simulations, are determined solely by the $F_x$ value of the sample. Moreover, we found that crack propagation exhibits very similar behaviors in MAC and maBN. Crack propagation can be regarded as the formation and the coalescence of voids. The existence of crystallites affects the locations of void formation, causing behaviors such as deflection, stopping, and bridging of cracks, which lead to rich toughening mechanisms compared with the crystalline material. The results deepen our understanding of the structure-mechanical-properties relationship in 2D amorphous materials.

**Results and discussion**

The MLP set is trained by using the open-source DeepMD-kit package [25, 26], and the root-mean-square-error (RMSE) of the validation set is about 6 meV per atom, which reached the widely accepted standard to give accurate MLPs. Details of the generation process are described in Supplemental Material [SM]. We first validated the reliability of the as-generated MLPs for kMC simulations. The energies of 100 accepted (red triangles) and 100 denied (blue inverted triangles) monolayer carbon and BN structures from kMC simulations with distinct DOD are compared in Figures 1a and 1b, respectively. The energies calculated by MLPs are very close to those calculated by DFT. Furthermore, for the formation energies of some typical defects in graphene and h-BN that affect kMC simulations, the MLP results are in agreement with those of DFT calculations (Table. S3, S4). To validate the stretching simulations, we stretched several samples, and collected 150 stretched and fractured structures of MAC and maBN. The energies of these structures (green dots) are also compared in Fig. 1a and 1b respectively. It is clear that the MLPs can describe stretched and fractured amorphous structures with DFT-level accuracy.

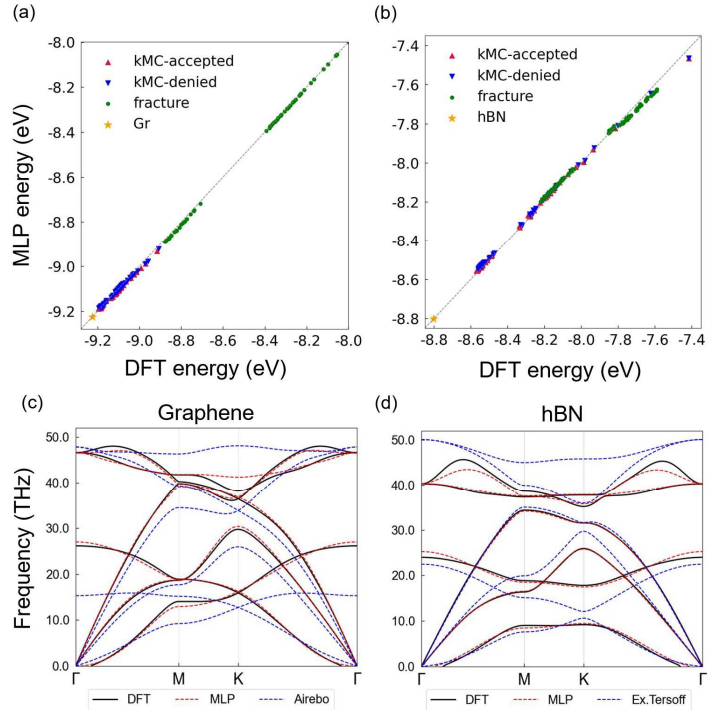

**Figure 1. Validation of MLP.** (a, b) Energy deviations between DFT and MLP calculations for rejected and accepted structures in kMC simulations, amorphous fractured structures, and crystalline structures of (a) carbon and (b) BN systems. (c, d) Phonon dispersion comparisons for (c) graphene and (d) h-BN, calculated using DFT (black solid), MLP (red dashed), and a state-of-the-art empirical potential (blue dashed).

We also provide a benchmark of MLPs in phonon dispersions, which are relevant to the mechanical properties. As shown in Fig. 1c and 1d, the calculated phonon dispersions of crystalline graphene and monolayer h-BN, using the as-generated MLPs, are in excellent agreement with the DFT results. In contrast, the calculated phonon dispersions using best-of-breed empirical potentials (AIREBO [27] for graphene, Extended Tersoff [28] for h-BN) show significant deviations from the DFT results. At the same time, the elastic constants and modulus of crystalline graphene and monolayer h-BN calculated by MLPs also outperform empirical potentials and are in good agreement with DFT results (Tables S1, S2). Overall, by comparing MLPs with DFT side by side, it can be concluded that the as-generated MLPs can describe crystalline, amorphous, stretched, and fractured systems, with DFT-level accuracies that the best-of-breed empirical potentials cannot match. MLPs are slowly becoming the standard for "DFT-level" simulations and calculations for systems that cannot be handled by straight DFT calculations.

The first step is to construct a reliable atomic structure using MLPs. We performed kMC simulations [24] of the structural evolution of monolayer amorphous materials (MAC and maBN). Starting from an initial configuration with randomly distributed atoms in a plane, five typical atomic structures of monolayer carbon and BN from different kMC steps are shown in Figs. 2a and 2b, respectively. The canonical hexagons in crystallite islands are colored green, while noncanonical hexagons in monolayer BN are colored blue. It is worth noting that, like monolayer carbon, monolayer BN also exhibits continuously growing crystallite regions during the kMC simulation with MLP. This result contrasts with earlier findings, based on kMC simulations using an empirical potential, that monolayer amorphous BN develops exclusively pseudocrystallites, namely honeycomb regions made up of noncanonical bonds [20]. We have now discovered that this difference arises because the extended Tersoff empirical potential substantially underestimates the energy of some noncanonical hexagons like those occurring in the recently predicted orthorhombic polymorph of BN (o-$B_2N_2$) [29], which directly affects the results of kMC simulations. In Table S4, we show that, unlike the empirical potential, the DFT-based MLPs reproduce the DFT-calculated formation energy of o-$B_2N_2$ very accurately. Thus, even though maBN contains two different elements and has a high possibility of forming noncanonical hexagons from a random distribution of atoms, the more stable, lower-energy crystallite structures prevail.

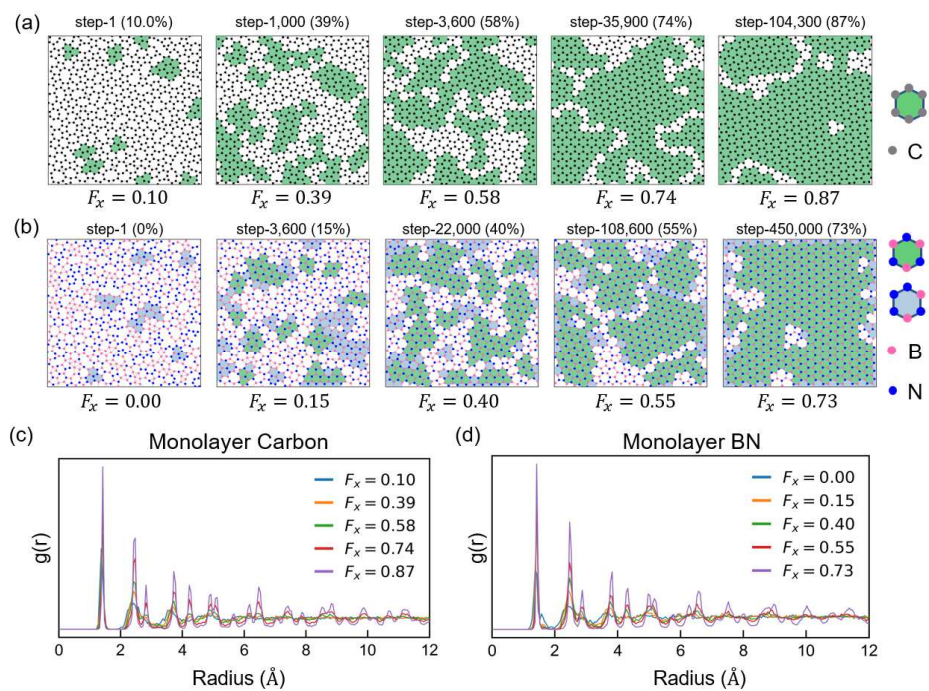

**Figure 2. Atomic structures of monolayer carbon and BN in kMC simulation.** (a-b) Atomic structures of monolayer carbon and BN from different kMC steps, respectively. The hexagons in crystallite regions are colored in green (canonical hexagons) and blue (noncanonical hexagons). The percentage of canonical hexagons and corresponding DOD order parameter $F_x$ are listed. (c-d) RDF of monolayer carbon and BN with five different $F_x$, respectively.

In order to distinguish amorphous structures with different DOD, an order parameter is necessary. Previously, Tian *et al.* defined a DOD order parameter, $\eta_{MRO}$, namely the ratio of the medium-range order (MRO) of amorphous and crystalline samples, obtained from the fluctuations of the experimental radial distribution functions (RDFs) in the medium-distance range. However, in order to correlate the conductivity with the DOD, it was found necessary to introduce a second order parameter, the density of conducting sites $\rho_{sites}$, which was derived directly from atomic-scale images of MAC samples.

We explored the applicability of $\eta_{MRO}$ to characterize the wide range of kMC-generated samples, i.e., individual kMC snapshots, of which Fig. 2a shows only five. We used MD simulations of several samples at room temperature and calculated their RDFs and $\eta_{MRO}$. We found that samples with very similar MRO and hence similar $\eta_{MRO}$ may differ significantly in their short-range RDFs, their bond-angle and bond-length distributions, and even more conspicuously in the fractions of the areas occupied by crystallites (see Fig. S3 for a detailed discussion). These observations motivated us to propose an alternative and more intuitive order parameter for monolayer amorphous materials that is directly based on the atomic structure, namely the fraction of the crystallite part of the structure, defined by

$$F_x = \frac{N_x}{N_{CRN}+N_x}.$$

Here $N_x$ and $N_{CRN}$ are the numbers of rings in the crystallites and the CRN regions, respectively (see more details in Supplemental Material). As a result, $F_x$ ranges from 0 for the most disordered structure (fully CRN) to 1 for the most ordered structures (crystalline graphene or maBN). The calculated $F_x$ values for samples with the same $\eta_{MRO}$ are quite distinct (see Table S5). The $F_x$ values of the five monolayer carbon samples in Fig. 2a are used to label their structures in Fig. 2a while their RDFs are shown in Fig. 2c. All structures show clear short-range order, but their RDF peaks are broadened differently. Structures with smaller $F_x$ exhibit broader RDF peaks and broader bond-angle and bond-length distributions, i.e., larger DOD (see Figs. S4 and S5). The five $F_x$ values are compared with the respective $\eta_{MRO}$ values in Table S6. The net conclusion is that $\eta_{MRO}$ appears to be relatively insensitive to increasing DOD in samples with CRN areas that occupy more than ~50% of the sample, i.e., for $F_x < 0.5$.

To validate the effectiveness of $F_x$, we generated samples with completely different sizes and distributions of crystallites, but with roughly equal values of $F_x$. The atomic structures, RDFs, and distributions of bond lengths and bond angles of three MAC samples with identical $F_x = 0.5$ are compared in Fig. 3. Similar comparisons are made for two more groups, each with three MAC samples, and $F_x = 0.25$ and $F_x = 0.75$, respectively, in Fig. S4. In all cases, we found that *samples with very different atomic structures but the same value of $F_x$, i.e., samples with very different sizes and arrangements of crystallites but similar total crystallite areal fraction, exhibit nearly identical RDFs and distributions of bond lengths and angles*. In other words, $F_x$ captures the essential indicators of DOD.

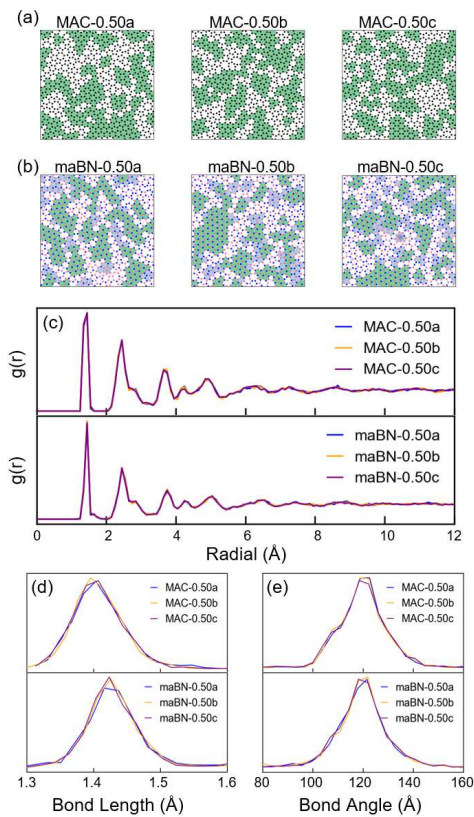

**Figure 3. Relation of $F_x$ to different manifestations of the DOD.** (a, b) Atomic structures of three different MAC and maBN samples with identical $F_x = 0.50$, distinguished by labels: a, b, and c. (c) RDFs of three MAC (upper panel) and maBN (lower panel) samples in (a, b). (d) Bond-length and (e) bond-angle distributions of the three MAC (upper panel) and three maBN (lower panel) samples in (a, b).

For the binary maBN systems, the values of $F_x$ are also calculated and shown in Fig. 2b while their RDFs are shown in Fig. 2d. We also generated samples with completely different local atomic structures, but with roughly equal values of $F_x$, and compared their RDFs and their distributions of bond lengths and bond angles, shown in Fig. 3 and Fig. S5. Once more, we find that *samples with very different atomic structures but the same value of $F_x$ exhibit nearly identical RDFs and distributions of bond lengths and bond angles*. As all the noncanonical B-B and N-N bonds are distributed in CRN regions, the areal fraction of crystallites, $F_x$, is able to capture both the structural and chemical DOD in binary monolayers. We have, therefore, established $F_x$ as an effective indicator (order parameter) of the DOD for 2D amorphous materials.

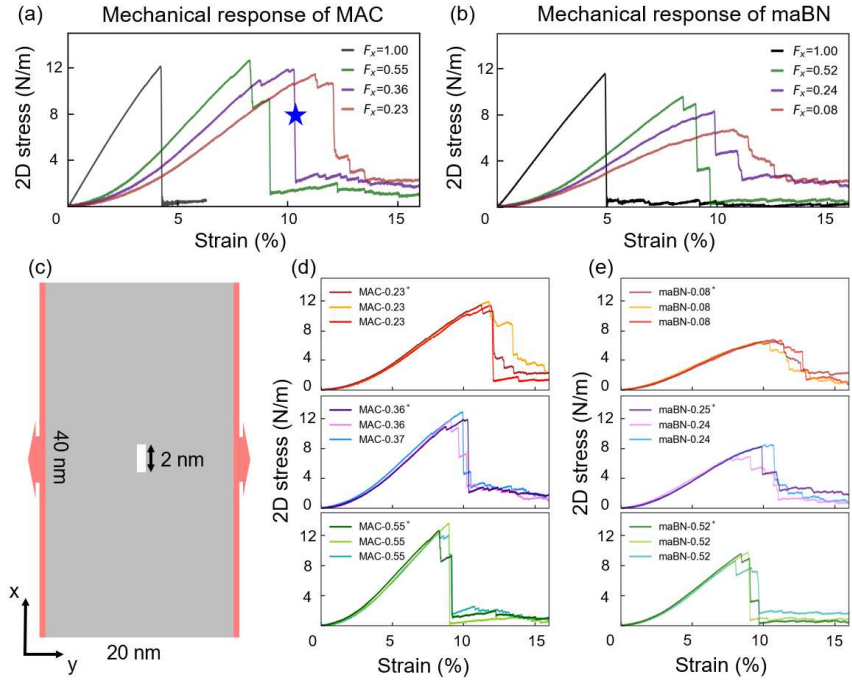

**Figure 4. Mechanical response of MAC and maBN.** (a) Strain-stress curves of MAC and with three different $F_x$ comparing with the precrack graphene along a zigzag direction. (b) Stress-strain curves of maBN with different $F_x$ compared with the precrack monolayer h-BN along a zigzag direction. (c) Schematic of stretching simulations. (d, e) Strain-stress curves for three groups of MAC and maBN samples, respectively. Samples in each group have similar $F_x$ values, but completely different atomic structures, with the starred sample in the legend corresponding to the sample in (a) for MAC and (b) for maBN.

Admittedly, the definition of $F_x$ as a measure of the DOD is accessible to experiments only through atomic-resolution images. However, this definition enables a detailed theoretical investigation of structure-properties relations. We have investigated the mechanical properties of MAC and maBN by performing MD simulations as follows. Structures with areas ~40×20 nm² were generated using the modified-building-blocks method [30]. Details are described in Fig. S6 and Fig. S7. As shown schematically in Fig. 4c, a 2-nm-long precrack was introduced at the center of the model. Then a far-field tensile load is applied in the *y*-direction until the sample breaks (a fracture develops throughout the sample). Figure 4a compares the nominal 2D stress-strain relations for the mechanical response of graphene along the zigzag direction and that of MAC with different values of $F_x$. In the small-strain region, the stress increases smoothly as the strain is enhanced. There are two differences between crystalline and amorphous mechanical responses: 1) the stress in MAC is much lower than that of graphene at the same strain; 2) unlike the linear stress-strain curve of graphene, the stress-strain curves of all MAC samples are nonlinear. This nonlinear stress-strain region suggests a plastic deformation in MAC samples, which is attributed to the rough surface and the relaxations of defects in MAC [22].

In a fracture process, when the strain increases to a critical value, the stress reaches its maximum and then drops, accompanied by the fracture propagation and the release of strain energy. As a typical brittle 2D material, graphene exhibits an abrupt drop in its stress-strain curve, while the stress of the MAC samples drops in a staircase fashion. Moreover, the curves in Fig. 4a show a sequential pattern: as the DOD increases, i.e., as $F_x$ decreases, the critical strain increases and the maximum stress decreases. We are not ready, however, to correlate $F_x$ with mechanical properties, because a given value of $F_x$ can correspond to different samples with distinct atomic structures. We, therefore, generated additional samples with similar values of $F_x$ but different atomic structures. Their stress-strain curves are shown in Fig. 4d. The stress-strain curves of different samples with similar $F_x$ are very similar to each other when the strain is smaller than the critical strain. Moreover, different structures with similar values of $F_x$ exhibit similar critical strain and maximum stress. Therefore, our results demonstrate that *the mechanical properties, i.e., the critical strain and the maximum stress of MAC is determined solely by $F_x$*.

As mentioned above, the stress-strain curves of MAC exhibit kinks with abrupt drops in stress (Fig. 4a). Each kink observed on the stress-strain curve indicates an initiation, propagation, or arrest of a crack. To analyze the crack behaviors near the kinks, the stress distribution of a snapshot of MAC ($F_x$ = 0.36) corresponding to the star mark on the stress-strain curve after the appearance of several kinks in Fig. 4a is shown in Fig. S8. It is found that there is no stress concentration near the crack tips of crack A and B, which suggests that cracks A and B already stopped [31]. These simulation results are consistent with experimental data on plasticity, large toughness, and arrested crack propagation in MAC [17].

We next turn to the mechanical properties of maBN. The stress-strain curves of h-BN along the zigzag direction and maBN samples with different and similar values of $F_x$ are shown in Fig. 4b and Fig. 4e. We see the maBN shares features with MAC, including the nonlinear stress-strain relation at the small-strain region and smaller critical strain at higher values of $F_x$. Note that, while graphene and h-BN exhibit distinct fracture properties — with graphene having an atomically smooth cracked edge [5] but h-BN lacking such smoothness [10] — they still demonstrate similar features after amorphization. Based on the present results shown in Fig. 4(b, e), we propose that the introduction of amorphousness may serve as a universal route to toughness enhancement of 2D materials.

Another obvious difference observed in the stress-strain curves of crystalline and amorphous materials is how stress decreases during the fracture process, which is associated with the propagation of cracks. In contrast to crystalline materials, stress does not immediately drop to zero in amorphous materials. Instead, the stress in MAC and maBN decreases slowly and even fluctuates. To understand how the crack propagates in MAC and maBN, a detailed analysis of crack propagation was performed.

Figure 5a shows the pathway of the main crack in one maBN sample. It is found that the crack propagates mainly through the CRN regions between crystallites. The same is also true for MAC. The randomly distributed crystallites embedded in the CRN regions result in a meandering crack path, which costs much more energy than a straight crack path like that of crystalline graphene [5]. The crack propagation in the CRN region can be regarded as the formation and coalescence of voids as shown in Fig. S9, which is similar to bulk amorphous carbon [32]. Stress concentrations are more likely to occur in CRN regions than in crystallite regions due to the existence of holes and inhomogeneities in CRN regions. The stress concentration near the crack tip leads to the formation of voids near the crack tip, and the crack tip extends towards the voids and connects with these voids to form the new tip.

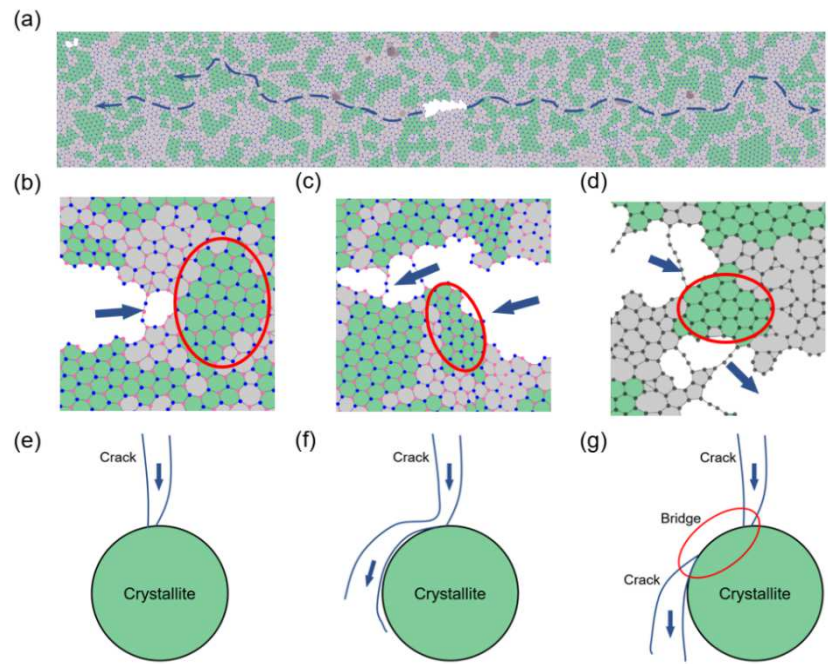

**Figure 5. Crack propagation.** (a) Crack path in crystallite-maBN. The main crack interacts with crystallites in three main ways: (b, e) crystallites stop the propagation of the main crack; (c, f) the main crack is deflected by crystallites; (d, g) while the main crack is stopped by crystallite, another crack initiates near the crystallite and a bridge is formed between the two cracks.

We next analyze the influence of crystallites on the crack propagation path. Some typical snapshots during crack propagation and corresponding schematics are shown in Fig. 5b-d. Figure 5b show one blunted crack tip (blue arrows mark the direction of crack propagation) pointing to a crystallite (labelled by the red circle). In addition, there is no stress concentration (Fig. S10) nor emergent voids near the crack tip, which indicates that the propagation of the crack stops at the crystallite. In contrast, voids continuously form near the crack tip shown in Fig. 5c. As a result, this crack changes its direction and propagates along the edge of a crystallite instead of being blunted by the crystallite. In other cases, we found two cracks near a crystallite such as the snapshot shown in Fig. 5d. One of the cracks stops in front of the crystallite just as the crack shown in Fig. 5b. The other crack initiates from a void far from the first crack tip. The two cracks are separated by the crystallite in between, which acts as a bridge between them.

Through a detailed analysis of these snapshots, the propagation of cracks can be understood as follows. Voids are the precursors of crack tips. However, voids are very difficult to form thermally in crystallites because the formation energies of vacancies are large (7.5 eV for the vacancy in graphene) [33]. When the crack tip reaches a crystallite, there are several possible outcomes. If there is no 'unstable structure' that is prone to cause stress concentration on either side of the crack, the degree of stress concentration at the crack tip is not enough to induce the formation of voids, whereby the propagation of the crack is forced to stop (Fig. 5b, 5e). On the contrary, if there is an 'unstable structure' nearby, where voids can form from, the crack is deflected by the crystallite (Fig. 5c, 5f). The deflection of the crack path increases the energy cost of crack propagation and hence toughens the materials.

Another possibility is that, after a propagating crack stops at a crystallite, voids form on the other side of the crystallite far from the crack tip, whereby they are not able to directly merge into the crack tip. In this case, a new crack tip initiates at such a void and the crystallite forms a bridge between the two cracks (Fig. 5d, 5g). The formation of the bridge helps reduce the local stress in the wake of the crack and toughens the material [34]. Overall, embedded crystallites terminate or deflect a crack that is propagating towards them or make the crack discontinuous. A meandering crack path increases the energy cost of crack propagation and leads to toughening of amorphous materials.

**Conclusions**

In summary, accurate MLPs for monolayer amorphous carbon and BN are trained with comprehensive sampling in the phase space by kMC-assisted active-learning. Crystallites are much more energetically favorable and can easily form in maBN, suggesting that pseudocrystallites are not likely to form in non-elemental materials. An intuitive order parameter, $F_x$, based on the atomic structures, is proposed to quantify the DOD in amorphous materials that comprise a Z-CRN and crystallites. Its effectiveness is demonstrated in MAC and maBN. For mechanical properties, large-scale uniform MAC and maBN samples were generated using the modified-building-block method. We find that the mechanical response before fracture, e.g., critical strain and stress, is solely determined by $F_x$. With the increase of $F_x$, there is a noticeable downward (upward) trend in critical strain (stress), respectively (exploration of how electrical conductivity of amorphous monolayers [18] correlates with $F_x$, however, is beyond the scope of this paper). A high crack resistance is observed in amorphous samples during the fracture process. Our analysis of crack propagations reveals that the crack resistance is attributed to complicated crack behaviors resulting from the presence of the crystallites in a Z-CRN amorphous structure. In disordered CRN regions, there is high stress concentration, leading to the formation of voids and crack propagation. Conversely, the crystallite regions, which possess resistance to void formation, can stop or deflect crack propagation or even induce the initiation of another crack, acting as a bridge between cracks. In both MAC and maBN, these behaviors are common and contribute to the propagation of cracks in a meandering and more energetically costly manner, which indicates that amorphization can toughen the two different materials in the same way. This finding suggests that amorphization may be a universal toughening mechanism, capable of improving the mechanical properties of various 2D materials.

**Acknowledgements**

This work was supported by the National Key R&D program of China (No. 2019YFA0308500), the National Natural Science Foundation of China (No. 52250402 and 61888102), CAS Project for Young Scientists in Basic Research (YSBR-003), and the Fundamental Research Funds for the Central Universities. A portion of the research was performed in CAS Key Laboratory of Vacuum Physics. Work at Vanderbilt was supported by the Department of Energy, Office of Science, Basic Energy Sciences, Materials Science and Engineering Division grant No. DE-FG02-09ER46554 and by the McMinn Endowment at Vanderbilt University.

**References:**

[1] Y. Kim, J. Lee, M. S. Yeom, J. W. Shin, H. Kim, Y. Cui, J. W. Kysar, J. Hone, Y. Jung, S. Jeon *et al.*, Nat. Commun. **4**, 2114 (2013).
[2] C. Lee, X. Wei, J. W. Kysar, J. Hone, Science **321**, 385 (2008).
[3] S. Kim, J. Yu, A. M. van der Zande, Nano Lett. **18**, 6686 (2018).
[4] B. Ni, D. Steinbach, Z. Yang, A. Lew, B. Zhang, Q. Fang, M. J. Buehler, J. Lou, MRS Bull. **47**, 848 (2022).
[5] P. Zhang, L. Ma, F. Fan, Z. Zeng, C. Peng, P. E. Loya, Z. Liu, Y. Gong, J. Zhang, X. Zhang *et al.*, Nat. Commun. **5**, 3782 (2014).
[6] T. Zhang, H. Gao, J. Appl. Mech. **82**, 051001 (2015).
[7] A. Shekhawat, R. O. Ritchie, Nat. Commun. **7**, 10546 (2016).
[8] Y. Yang, X. Li, M. Wen, E. Hacopian, W. Chen, Y. Gong, J. Zhang, B. Li, W. Zhou, P. M. Ajayan *et al.*, Adv. Mater. **29**, 1604201 (2017).
[9] G. Lopez-Polin, J. Gomez-Herrero, C. Gomez-Navarro, Nano Lett. **15**, 2050 (2015).
[10] Y. Yang, Z. Song, G. Lu, Q. Zhang, B. Zhang, B. Ni, C. Wang, X. Li, L. Gu, X. Xie *et al.*, Nature **594**, 57 (2021).
[11] J. Robertson, Phys. Rev. Lett. **68**, 220 (1992).
[12] C. Fan, C. Li, A. Inoue, V. Haas, Phys. Rev. B **61**, R3761 (2000).
[13] Z. P. Lu, C. T. Liu, J. R. Thompson, W. D. Porter, Phys. Rev. Lett. **92**, 245503 (2004).
[14] V. I. Ivashchenko, P. E. A. Turchi, V. I. Shevchenko, Phys. Rev. B **75**, 085209 (2007).
[15] C. A. Schuh, T. C. Hufnagel, U. Ramamurty, Acta Mater. **55**, 4067 (2007).
[16] M. Zhu, J. Zhou, Z. He, Y. Zhang, H. Wu, J. Chen, Y. Zhu, Y. Hou, H. Wu, Y. Lu, Mater. Horiz. (2023).
[17] C. T. Toh, H. Zhang, J. Lin, A. S. Mayorov, Y. P. Wang, C. M. Orofeo, D. B. Ferry, H. Andersen, N. Kakenov, Z. Guo *et al.*, Nature **577**, 199 (2020).
[18] H. Tian, Y. Ma, Z. Li, M. Cheng, S. Ning, E. Han, M. Xu, P. F. Zhang, K. Zhao, R. Li *et al.*, Nature **615**, 56 (2023).
[19] S. Hong, C. S. Lee, M. H. Lee, Y. Lee, K. Y. Ma, G. Kim, S. I. Yoon, K. Ihm, K. J. Kim, T. J. Shin *et al.*, Nature **582**, 511 (2020).
[20] Y. T. Zhang, Y. P. Wang, X. Zhang, Y. Y. Zhang, S. Du, S. T. Pantelides, Nano Lett. **22**, 8018 (2022).
[21] L. C. Felix, R. M. Tromer, P. A. S. Autreto, L. A. Ribeiro Junior, D. S. Galvao, J. Phys. Chem. C **124**, 14855 (2020).
[22] W. Xie, Y. Wei, Nano Lett. **21**, 4823 (2021).
[23] Y.-T. Zhang, Y.-P. Wang, Y.-Y. Zhang, S. Du, S. T. Pantelides, Appl. Phys. Lett. **120**, 222201 (2022).
[24] F. Ding, B. I. Yakobson, J. Phys. Chem. Lett. **5**, 2922 (2014).
[25] H. Wang, L. Zhang, J. Han, W. E, Comput. Phys. Commun. **228**, 178 (2018).
[26] L. Zhang, J. Han, H. Wang, R. Car, W. E, Phys. Rev. Lett. **120**, 143001 (2018).
[27] S. J. Stuart, A. B. Tutein, J. A. Harrison, J. Chem. Phys. **112**, 6472 (2000).
[28] J. H. Los, J. M. H. Kroes, K. Albe, R. M. Gordillo, M. I. Katsnelson, A. Fasolino, Phys. Rev. B **96**, 184108 (2017).
[29] S. Demirci, S. E. Rad, S. Kazak, S. Nezir, S. Jahangirov, Phys. Rev. B **101**, 125408 (2020).
[30] B. Cai, X. Zhang, D. A. Drabold, Phys. Rev. B **83**, 092202 (2011).
[31] A. A. Griffith, Philos. Trans. R. Soc. London, A **221**, 163 (1921).
[32] S. M. Khosrownejad, J. R. Kermode, L. Pastewka, Phys. Rev. Mater. **5**, 023602 (2021).

[33] M. D. Bhatt, H. Kim, G. Kim, RSC Adv. **12**, 21520 (2022).
[34] R. O. Ritchie, Nat. Mater. **10**, 817 (2011).

# Supplemental Material for

# "Structure and mechanical properties of monolayer amorphous carbon and boron nitride"

Xi Zhang[1#], Yu-Tian Zhang[1,2#], Yun-Peng Wang[3], Shiyu Li[1], Shixuan Du[1,4], Yu-Yang Zhang[1*], Sokrates T. Pantelides[5,1]

1. University of Chinese Academy of Sciences and Institute of Physics, Chinese Academy of Sciences, Beijing 100049, China

2. CAS Key Laboratory of Theoretical Physics, Institute of Theoretical Physics, Chinese Academy of Sciences, Beijing 100190, China

3. Hunan Key Laboratory for Super Microstructure and Ultrafast Process, School of Physics and Electronics, Central South University, Changsha 410083, China

4. Songshan Lake Materials Laboratory, Dongguan, Guangdong 523808, China

5. Department of Physics and Astronomy and Department of Electrical and Computer Engineering, Vanderbilt University, Nashville, Tennessee 37235, US

*Email: zhangyuyang@ucas.ac.cn

## Kinetic-Monte-Carlo (kMC) simulations

Kinetic-Monte-Carlo (kMC) [1, 2] simulations are employed to obtain atomic configurations of MAC and maBN with different DOD. The kMC simulation has been widely used in simulating the dynamical annealing process [1-4].

For both MAC and maBN, the atomic number density of structures is the same as in crystalline graphene and monolayer h-BN, respectively, and the B:N ratio is 1:1. During kMC simulations, two neighboring atoms are randomly selected for a Stone-Wales (SW) transformation at each step. Considering the binary nature of BN, both SW and anti-site transformations (exchange two atoms) are considered. Then the structures are relaxed and accepted by a probability defined as min{1, exp[-($E_{new}$-$E_{old}$)/$k_B$T]}, where $E_{old}$ is the energy of the current configuration, $E_{new}$ is the energy of the new configuration, and $k_B$T is set to 0.5 eV.

## Molecular dynamics (MD) simulations

MD simulations are performed using the large-scale atomic/molecular, massively parallel simulator (LAMMPS) [5] with the MLPs. Mechanical responses were evaluated by conducting a uniaxial tensile simulation with a time increment of 1 fs.

Before applying the loading conditions, all structures were equilibrated using the Nosé–Hoover barostat and thermostat method (NPT) at 300K. Then, a constant engineering strain ($10^{-4}$ $ps^{-1}$) was applied, and the NVT ensemble is employed to control the temperature fluctuations. In order to compare the stress of different monolayers, we use the nominal 2D stress, $\sigma_{2D}=F/L$, where $L$ is the length of the side on which the force is applied (force per unit length).

## Machine Learning Potentials (MLPs)

Recently developed machine learning potentials (MLPs) [6, 7] bring us a powerful tool to perform accurate molecular dynamics simulations in complex systems. By fitting the energy, force, and stress from density-functional-theory (DFT) calculations, the MLPs can reach DFT-level accuracy. However, the quality of the MLPs mainly depends on the quality of the training data, which consists of a series of structures and corresponding total energy, atomic force, and stress calculated by a higher-level calculation such as DFT.

**Training (kMC-assisted active-learning):**

To ensure the performance of MLP in complex amorphous systems, the training data have to include structures as diverse as possible. In our work, kinetic-Monte-Carlo (kMC) [2] simulations are employed to explore the potential energy surface (PES) and collect structures with different degrees of disorder (DOD). The open-source active-learning package DPGEN [8] is used to refine the dataset.

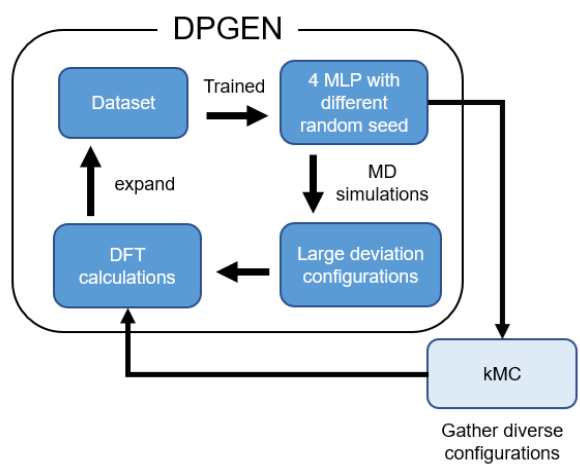

**Figure S1. Workflow of kMC-assisted active-learning.**

Figure S1 shows the workflow. We first perform some KMC simulations with empirical potentials and collect a series of 2D structures with different DOD, including continuous random network (CRN), crystallite, pseudocrystallite, nanocrystalline, polycrystalline, and crystalline as the initial dataset for training the MLP, which in turns helps the KMC simulation with data collecting (Fig. S1). For fracture situations, some stretching simulations are performed to collect the fracture structures. The DPGEN is employed in the entire procedure to expand and refine the dataset.

In the dataset, the energy, force, and virial stress of all the configurations are calculated using the Vienna Ab initio Simulation Package (VASP v6.3.2) [9]. A plane wave cutoff of 500 eV, the Perdew–Burke–Ernzerhof (PBE) exchange-correlation functional [10], is employed. The PBE functional is considered suitable for the present investigations because they are based entirely based on total energies of ground-state electronic configurations (hybrid functionals are needed when excited states, e.g., electronic band gaps, are calculated). The K-grid is auto-generated by KSPACING=0.333.

The individual active-learning datasets are combined to train the final potential. The smooth edition of DeePMD, DeePot-SE model (se_e2_a)[11] as implemented in the DeePMD-kit package [12, 13] was used to train the MLP. The cutoff radius of the model was set as 6.5Å for neighbor searching while the smoothing function decays from 4.0Å. The sizes of hidden layers of embedding net from the input end to the output end are 25, 50, and 100, respectively. The fitting net consists of three hidden layers with 120 neurons in each layer. Hyperbolic Tangent was employed as an activation function. The learning rate decayed from $10^{-3}$ to $3.51 \times 10^{-8}$ exponentially. 90% of the dataset was randomly selected for training and the rest for validation.

**Validation:**

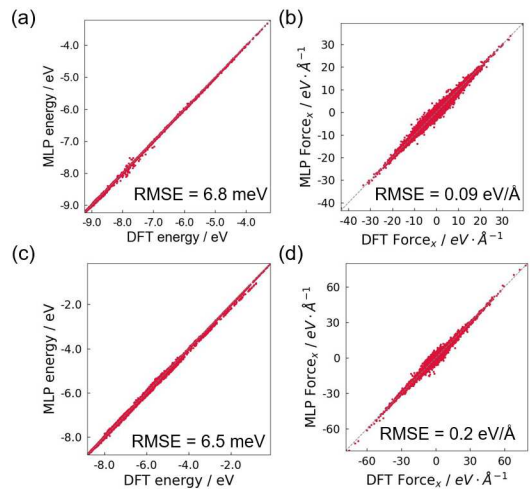

**Figure S2. Deviation of our MLP in validation dataset** (a) energy of carbon systems; (b) force of carbon system; (c) energy of BN systems; (b) force of BN system.

The deviations of energy and force between MLPs and DFT calculations are shown in Figure S2. The root-mean-square-error (RMSE) of the validation dataset is 6.8 (6.5) meV and 0.09 (0.2) eV/Å for energy and force in the carbon (boron nitride) system, respectively.

Table S1 Elastic constants of graphene.

| (GPa) | DFT | MLP | AIREBO [14] |
|---|---|---|---|
| $C_{11}$ | 1058.647 | 1074.683 | 878.593 |
| $C_{12}$ | 182.545 | 208.180 | 283.736 |
| $C_{66}$ | 438.048 | 433.252 | 297.4311 |
| Modulus | 1026.985 | 1034.356 | 786.962 |

Table S2. Elastic modulus of h-BN.

| (GPa) | DFT | MLP | Ext. Tersoff [15] |
|---|---|---|---|
| $C_{11}$ | 869.611 | 862.018 | 832.305 |
| $C_{12}$ | 193.491 | 205.982 | 149.114 |
| $C_{66}$ | 338.060 | 328.018 | 341.599 |
| Modulus | 826.559 | 812.798 | 805.590 |

In Tables S1 and S2, we compare the elastic constants of graphene and h-BN calculated by DFT, MLPs, and the best-of-breed empirical potentials. In each material, the MLP are in good accord with the corresponding DFT results. The formation energies of some typical defects, including Stone-Wales (SW), single vacancy (SV), double vacancies (DV), triple vacancies (TV), and tetra vacancies (TeV), in graphene (Table S3) and h-BN (Table S4) are compared. It is found that both MLP and empirical potential values are consistent with DFT calculations. However, when we consider the orthorhombic boron nitride (o-$B_2N_2$) [16], a stable single-layer crystal structure of boron nitride that comprises a kind of non-canonical hexagons, the MLPs and DFT give very similar formation energies, whereas the Extended Tersoff give very different results. The same kind of non-canonical hexagons exist in large numbers in pseudocrystallite-maBN generated using the Extended Tersoff. Thus, the inability of the Extended Tersoff potential to describe o-$B_2N_2$ accurately results in the preference for pseudocrystallites in maBN [4]

Table S3. The formation energy of some typical defects in graphene calculated by different methods.

| Formation Energy (eV) | DFT | MLP | AIREBO |
|---|---|---|---|
| SW | 4.916 | 4.907 | 5.697 |
| SV | 8.021 | 6.042 | 7.928 |
| DV | 7.701 | 7.210 | 9.990 |
| TV | 11.243 | 10.230 | 13.409 |
| TeV | 10.923 | 11.426 | 14.719 |

Table S4. The formation energy of some typical defects in h-BN calculated by different methods.

| Formation Energy (eV) | DFT | MLP | Ext. Tersoff |
|---|---|---|---|
| o-$B_2N_2$[16] | 3.295 | 3.252 | -1.021 |
| SW | 6.949 | 7.434 | 9.693 |
| SV(B) | 7.290 | 6.737 | 12.003 |
| SV(N) | 8.342 | 5.784 | 7.093 |
| DV | 8.635 | 8.588 | 13.636 |
| TV | 10.768 | 10.334 | 19.118 |
| TeV | 11.142 | 11.684 | 24.238 |

# Order parameter for degree of disorder (DOD)

In a previous study [17], Tian *et al.* defined $\eta_{\text{MRO}}$, namely the level of medium-range order (MRO), as one of two order parameters that are needed to describe the DOD–conductivity relationship in MAC. It is defined by

$$\eta_{\text{MRO}} = \frac{A_{\text{MRO}}(a)}{A_{\text{MRO}}(c)},$$

where $A_{MRO}(a)$ and $A_{MRO}(c)$ are the areas of the "medium-range order" regions in the RDF (6 to 10 Å) of the amorphous and crystalline materials, respectively. This definition of a DOD order parameter is suitable for experimental investigations since it can be directly calculated from measured RDFs, as was done in Ref. [17]. However, in order to correlate the conductivity with the DOD, it was found necessary to introduce a second order parameter, the density of conducting sites $\rho_{sites}$, which was derived directly from atomic-scale images of MAC samples.

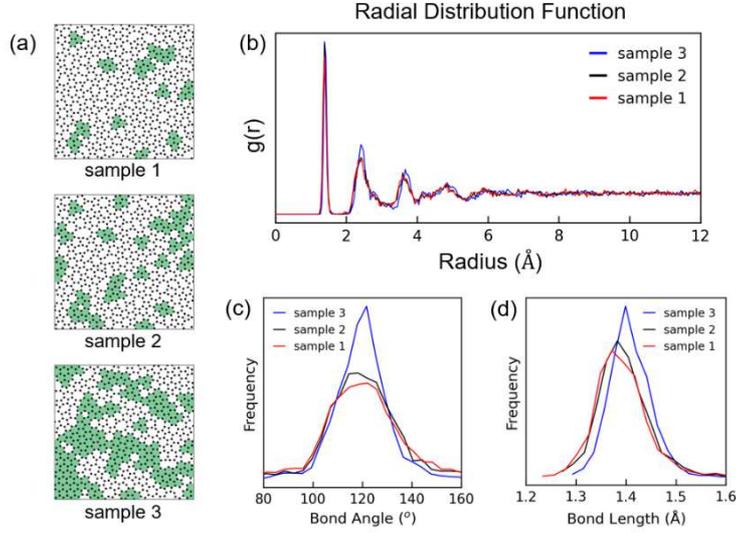

**Figure S3. Three different MAC samples with identical $\eta_{MRO}$ value of 0.03** (a) atomic structures of three different MAC samples; the areas covered by crystallites are distinctly different. (b) RDFs of the three MAC samples in (a); (c) bond-angle and (d) bond-length distributions of the three MAC samples in (a).

We investigated the applicability of $\eta_{MRO}$ to characterize the multitude of kMC amorphous structures (snapshots) and found that samples with similar $\eta_{MRO}$ may have very different DOD. Three such samples with the same MRO and hance the same $\eta_{MRO}$ (0.03) are shown in Fig. S3a. Their short-range RDFs as well as the bond-angle and bond-length distributions (Figs. S3b-d, respectively), all of which comprise part of the DOD, are different. An even more conspicuous difference in the DOD of the three samples is the fraction of the area occupied by crystallites.

The above observations, led us to define the alternative order parameter, $F_x$, as the fraction of crystallite regions in the whole material. The fraction is defined by the ratio of the number of rings in the crystallites and the number of all the rings in the entire sample. In practice, one determines the number of hexagons $N_x$ in all the crystallites in a sample and the number of rings in the CRN, $N_{CRN}$.) Then,

$$F_x = \frac{N_x}{N_{CRN}+N_x}.$$

The criterion for a hexagon to be part of a crystallite is that it must be attached to two other adjacent hexagons. In maBN, the crystallite hexagons must be canonical and be attached to two adjacent canonical hexagons. Other noncanonical hexagons are treated as parts of the Z-CRN. This definition of a DOD order parameter is suitable for theoretical investigations as it depends on the atomic structure of amorphous monolayers.

**Table S5 Comparison of DOD order parameters in samples with high DOD.**

| Sample | Formation Energy (eV/atom) | $F_x$ | $\eta_{MRO}$ |
|---|---|---|---|
| 1 | 0.53 | 0.11 | 0.03 |
| 2 | 0.44 | 0.25 | 0.03 |
| 3 | 0.28 | 0.46 | 0.03 |

The calculated $F_x$ values of the three MAC samples with identical $\eta_{MRO}$ in Fig. S3 are listed in Table S5 along with their $\eta_{MRO}$ and formation energy, defined by

$$E_f = E_{ave}(a) - E_{ave}(c),$$

where $E_{ave}(a)$ and $E_{ave}(c)$ are the energies per atom of the amorphous and crystalline samples, respectively. Their different DOD, exhibited in Fig. S3, are also reflected in their formation energies, which decrease from 0.53 to 0.28 eV/atom while their $F_x$ increase from 0.11 to 0.46. Thus, $F_x$ clearly distinguishes the DOD differences among samples that have similar $\eta_{MRO}$.

The order parameters $F_x$ and $\eta_{MRO}$ of graphene and the five MAC structures with completely different DOD depicted in Fig. 2a of the main text have been calculated and are listed in Table S6. The RDFs of these five samples are shown in Fig. 2c of the main text. The limited sensitivity of $\eta_{MRO}$ in characterizing materials of high degree of amorphousness can also be seen in the case of MAC-1 and MAC-2: the formation energy decreases from 0.62 to 0.31 eV/atom and $F_x$ increases from 0.10 to 0.39, while $\eta_{MRO}$ only changes from 0.04 to 0.03. Furthermore, $F_x$ is also effective and sensitive for samples with lower DOD as it smoothly drops from 1.00 for graphene to 0.74 for MAC-4 while $\eta_{MRO}$ drops precipitously from 1 to 0.14 and after that it decreases very slowly.

**Table S6 Comparison of DOD order parameters.**

| Sample | Formation Energy (eV/atom) | $F_x$ | $\eta_{MRO}$ |
|---|---|---|---|
| MAC-1 | 0.62 | 0.10 | 0.03 |
| MAC-2 | 0.31 | 0.39 | 0.04 |
| MAC-3 | 0.21 | 0.58 | 0.06 |
| MAC-4 | 0.13 | 0.74 | 0.14 |
| MAC-5 | 0.08 | 0.87 | 0.25 |
| Graphene | 0 | 1.00 | 1.00 |

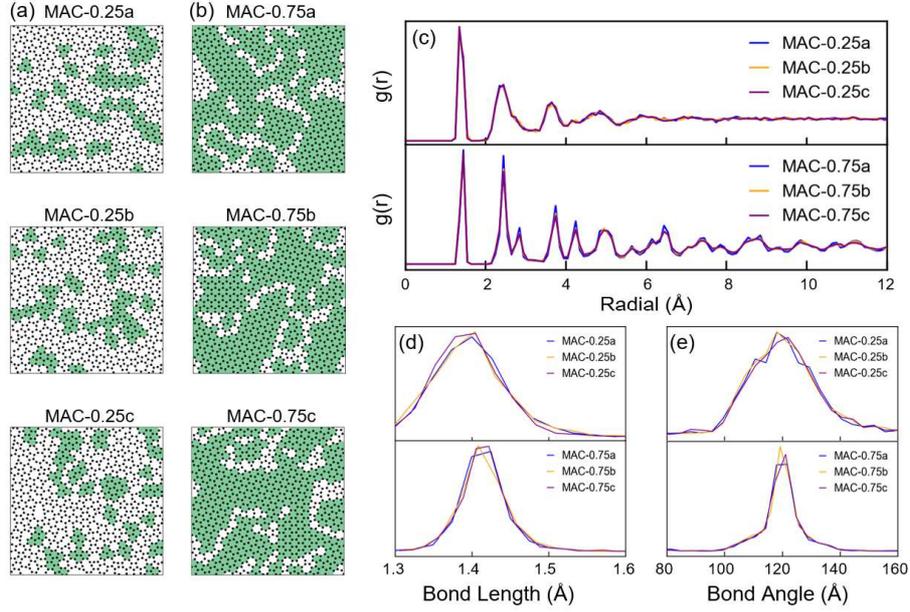

**Figure S4. Relation of $F_x$ to different manifestations of the MAC DOD.** (a-b) Atomic structures of two groups of MAC samples with identical $F_x$ value of 0.25 and 0.75, respectively. Each groups containing three different MAC samples, distinguished by labels: a, b, and c. (c) RDFs of two groups of MAC samples in (a, b) (in each group, the purple line covers the other two colors almost completely). (d) Bond length and (d) bond angle distribution of two groups MAC samples in (a, b).

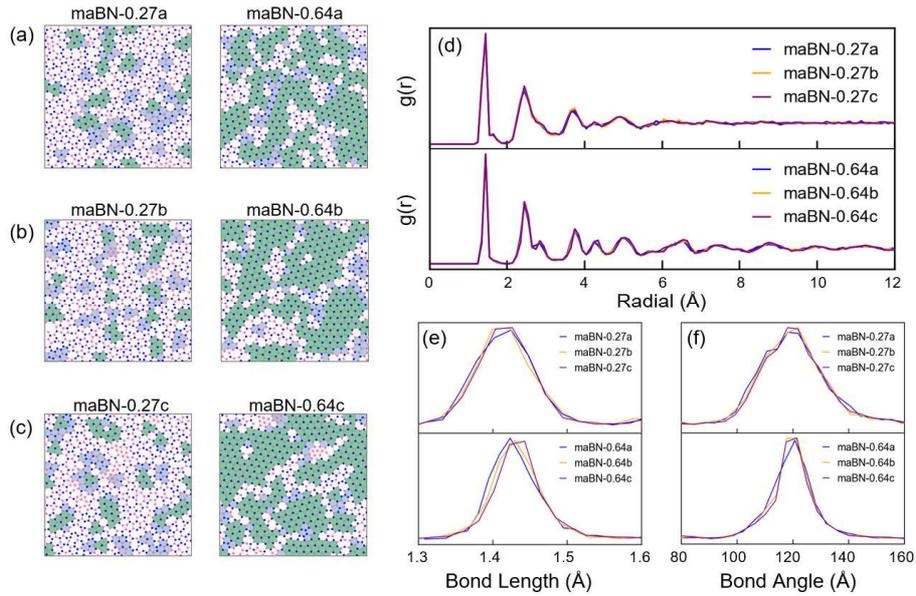

**Figure S5. Relation of $F_x$ to different manifestations of the maBN DOD.** (a-b) Atomic structures of two groups of maBN samples with identical $F_x$ value of 0.27 and 0.64, respectively. Each groups containing three different maBN samples, distinguished by labels: a, b, and c. (C) RDFs of two groups

of maBN samples in (a, b) (in each group, the purple line covers the other two colors almost completely). (d) Bond length and (d) bond angle distribution of two groups MAC samples in (a, b).

Since $F_x$ is defined as the fraction of crystallite region, we need to examine the relationship between $F_x$ and other structural information for validation. We collect three groups of MAC and three groups of maBN structures with roughly equal $F_x$ in each group, but with recognizably different atomic structures. The group of MAC and maBN with $F_x$ value of 0.5 is shown in Fig. 3 of the main text. The atomic structures, as well as RDFs and distributions of bond angles and bond lengths, of other two groups are shown in Figs. S4 and Fig. S5 for MAC and maBN, respectively. It is found that samples with the same $F_x$ exhibit nearly identical RDFs and distributions of bond lengths and bond angles, which indicates that they have the same DOD. Since the $\eta_{MRO}$ is determined by the RDF and samples with the same $F_x$ exhibit very similar RDFs, they also have similar $\eta_{MRO}$. As Table S5 demonstrates, however, MAC samples with very similar $\eta_{MRO}$ can have very different $F_x$. Overall, we conclude that $F_x$ is both intuitive and effective in capturing the DOD of MAC with sufficient sensitivity. Additionally, $F_x$ also serves well as an order parameter in maBN, the binary amorphous system.

## Modified Building blocks method:

The building blocks method [18], which is commonly used in amorphous systems, was employed to build large-scale monolayer amorphous structures for stretching simulations. In this work, we used kMC simulation instead of the melt-quench procedure to generate the amorphous structures, and we add some modifications to reduce the simulation times.

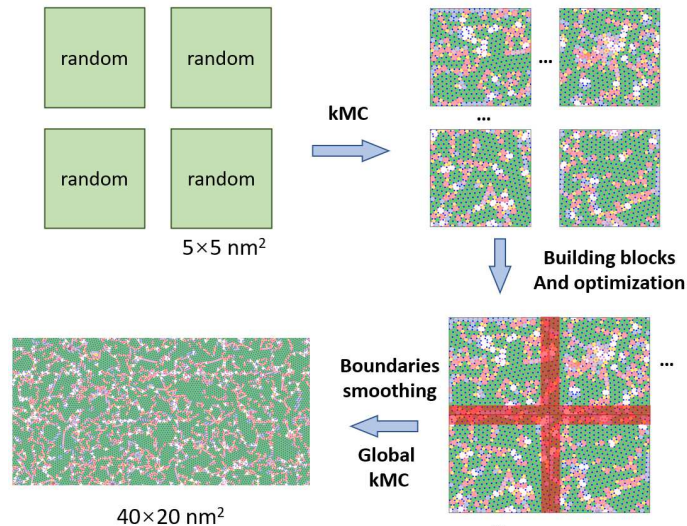

**Figure S6. Workflow of modified building blocks method for monolayer amorphous materials**

Figure S6 shows the workflow of the modified building blocks method. First, we generated 32 uncorrelated 5×5 nm² monolayer amorphous structures with a similar $F_x$. Then the 400×200 nm² sample is built from the building blocks. To smooth the boundaries that connect the building blocks (red region in Figure S6), we did the constraint kMC, in which we only did the SW transformation and bond exchange near the boundaries until the DOD of the boundaries and DOD of the building blocks are similar. Finally, we performed a global kMC in the entire structure to ensure uniformity.

**Validation of modified building block methods:**

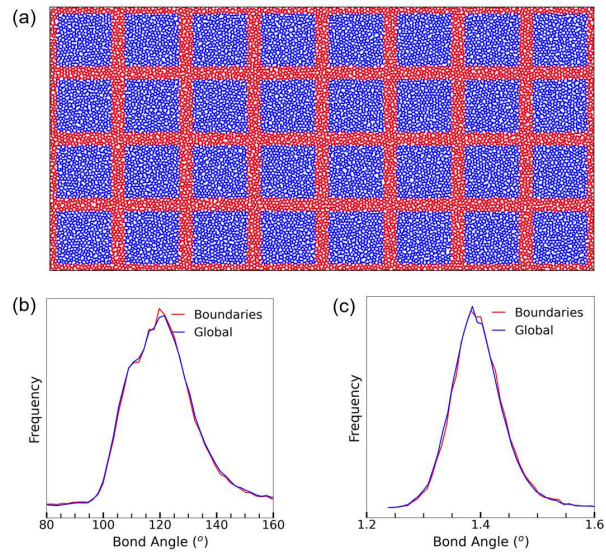

**Figure S7. Validation of constraint building blocks method.** (a) atomic structure of large-scale sample generated by building blocks methods, atoms in the boundary region are colored in red while other atoms are colored in blue, (b, c) distribution of bond angle and length in the boundary region and in the entire sample are compared

Figure S7a shows the atomic structure of a building block sample. The boundary regions and other regions are colored in red and blue respectively. To examine the validity of the modified building block method, the bond angles and lengths distribution of different regions are calculated and shown in Figure S7. It is found that the bond angles and lengths distribution of two different regions are very similar, which indicates that the entire sample is uniform and the modified building block method is valid.

# Additional simulation results:

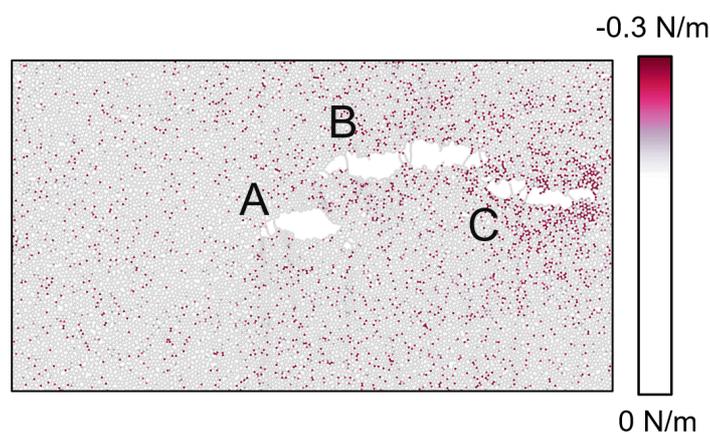

**Figure S8. Stress distribution of a MAC sample with $F_x$ =0.36 during the stretching simulation**

In order to demonstrate the arrested crack propagation, we selected a snapshot during the stretching simulation, which experienced two stress drops (two kinks on stress-strain curve). The stress distribution in this particular snapshot was calculated and illustrated in Fig. S8. This snapshot contains three cracks, among which cracks A and B exhibit no obvious stress concentration at their crack tips, while only the right-side crack tip of crack C shows stress concentration. This finding indicates that two of the cracks in the snapshot have undergone an arrested propagation, and their arrests correspond with two kinks on the stress-strain curve.

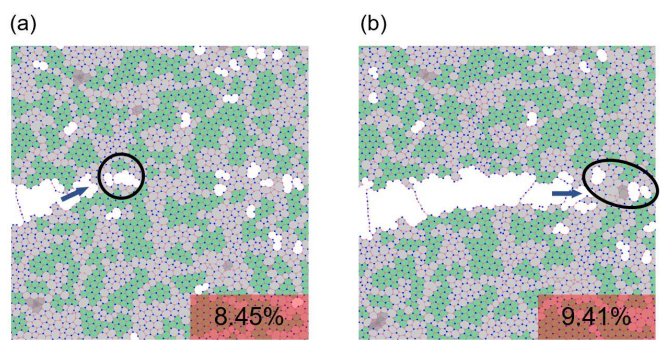

**Figure S9. Formation and coalescence of voids.**

Figure S9 shows two snapshots of maBN during crack propagation. The voids (labeled by the black circle) constantly form near the crack tip (labeled by the blue arrow) and coalesce with the crack. The voids form near some randomly distributed holes, which results in a meandering crack path.

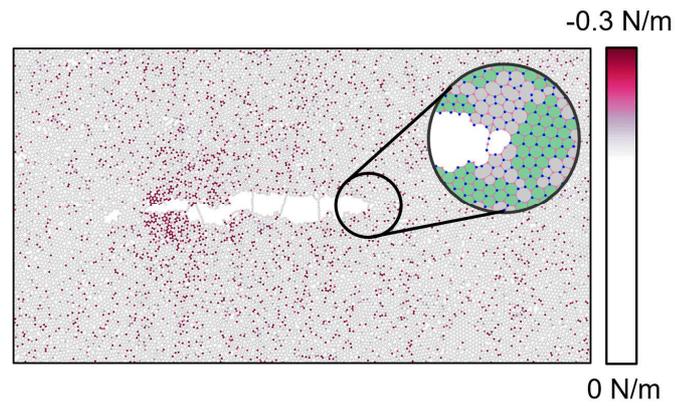

**Figure S10. Stress distribution of a maBN samples with the value of $F_x = 0.52$ during the stretching simulation**

  Figure S10 shows the stress distribution of a MAC sample a crack reaches a crystallite and stops propagating. The blunted crack tip is magnified to illustrate the atomic structure, with the crystallite (CRN) colored green (gray) and labeled by a black circle. Comparing to the propagating crack tip, it is found that there is no stress distribution near the blunted crack tip, which indicates the crack tip already stopped.

**References:**

[1] Y. Wei, J. Wu, H. Yin, X. Shi, R. Yang, M. Dresselhaus, Nat. Mater. **11**, 759 (2012).

[2] F. Ding, B. I. Yakobson, J. Phys. Chem. Lett. **5**, 2922 (2014).

[3] C. T. Toh, H. Zhang, J. Lin, A. S. Mayorov, Y. P. Wang, C. M. Orofeo, D. B. Ferry, H. Andersen, N. Kakenov, Z. Guo *et al.*, Nature **577**, 199 (2020).

[4] Y. T. Zhang, Y. P. Wang, X. Zhang, Y. Y. Zhang, S. Du, S. T. Pantelides, Nano Lett. **22**, 8018 (2022).

[5] S. Plimpton, J. Comput. Phys. **117**, 1 (1995).

[6] J. Behler, M. Parrinello, Phys. Rev. Lett. **98**, 146401 (2007).

[7] J. Behler, J. Chem. Phys. **134**, 074106 (2011).

[8] Y. Zhang, H. Wang, W. Chen, J. Zeng, L. Zhang, H. Wang, W. E, Comput. Phys. Commun. **253**, 107206 (2020).

[9] G. Kresse, J. Furthmüller, Comput. Mater. Sci. **6**, 15 (1996).

[10] J. P. Perdew, K. Burke, M. Ernzerhof, Phys. Rev. Lett. **77**, 3865 (1996).

[11] L. Zhang, J. Han, H. Wang, W. A. Saidi, R. Car, W. E, arXiv:1805.09003 (2018).

[12] H. Wang, L. Zhang, J. Han, W. E, Comput. Phys. Commun. **228**, 178 (2018).

[13] L. Zhang, J. Han, H. Wang, R. Car, W. E, Phys. Rev. Lett. **120**, 143001 (2018).

[14] S. J. Stuart, A. B. Tutein, J. A. Harrison, J. Chem. Phys. **112**, 6472 (2000).

[15] J. H. Los, J. M. H. Kroes, K. Albe, R. M. Gordillo, M. I. Katsnelson, A. Fasolino, Phys. Rev. B **96**, 184108 (2017).

[16] S. Demirci, S. E. Rad, S. Kazak, S. Nezir, S. Jahangirov, Phys. Rev. B **101**, 125408 (2020).

[17] H. Tian, Y. Ma, Z. Li, M. Cheng, S. Ning, E. Han, M. Xu, P. F. Zhang, K. Zhao, R. Li *et al.*, Nature **615**, 56 (2023).

[18] B. Cai, X. Zhang, D. A. Drabold, Phys. Rev. B **83**, 092202 (2011).